\documentclass[aps,amsmath,amssymb,reprint,superscriptaddress,longbibliography,UTF8]{revtex4-1}

\usepackage{float}
\usepackage{graphicx}
\usepackage{dcolumn}
\usepackage{bm}
\usepackage{mathrsfs,amssymb,amsmath}
\usepackage{ulem}
\usepackage[dvipsnames]{xcolor}
\usepackage[colorlinks=true,linkcolor=blue,citecolor=blue,anchorcolor=blue,urlcolor=blue]{hyperref}

\expandafter\ifx\csname package@font\endcsname\relax\else
 \expandafter\expandafter
 \expandafter\usepackage
 \expandafter\expandafter
 \expandafter{\csname package@font\endcsname}%
\fi
\begin{document}

\title{Calculating temperature-dependent properties of Nd$_2$Fe$_{14}$B permanent magnets by atomistic spin model simulations}

\author{Qihua Gong}
\affiliation{Institute of Materials Science, Technische Universit\"{a}t Darmstadt, 64287 Darmstadt, Germany}
\author{Min Yi}
\email{yimin@nuaa.edu.cn}
\affiliation{Institute of Materials Science, Technische Universit\"{a}t Darmstadt, 64287 Darmstadt, Germany}
\affiliation{State Key Lab of Mechanics and Control of Mechanical Structures \& Key Lab for Intelligent Nano Materials and Devices of Ministry of Education \& College of Aerospace Engineering, Nanjing University of Aeronautics and Astronautics (NUAA), Nanjing 210016, China}
\affiliation{State Key Lab for Strength and Vibration of Mechanical Structure, Xi'an Jiaotong University, Xi'an 710049, China}
\author{Richard F. L. Evans}
\affiliation{Department of Physics, The University of York, York, YO10 5DD, United Kingdom}
\author{Bai-Xiang Xu}
\affiliation{Institute of Materials Science, Technische Universit\"{a}t Darmstadt, 64287 Darmstadt, Germany}
\author{Oliver Gutfleisch}
\affiliation{Institute of Materials Science, Technische Universit\"{a}t Darmstadt, 64287 Darmstadt, Germany}

\date{\today}

\begin{abstract}
Temperature-dependent magnetic properties of Nd$_2$Fe$_{14}$B permanent magnets, i.e., saturation magnetization $M_\text{s}(T)$, effective magnetic anisotropy constants $K_i^\text{eff}(T)$ ($i=1,2,3$), domain wall width $\delta_w(T)$, and exchange stiffness constant $A_\text{e}(T)$, are calculated by using \textit{ab-initio} informed atomistic spin model simulations. 
We construct the atomistic spin model Hamiltonian for Nd$_2$Fe$_{14}$B by using the Heisenberg exchange of Fe$-$Fe and Fe$-$Nd atomic pairs, the uniaxial single-ion anisotropy of Fe atoms, and the crystal-field energy of Nd ions which is approximately expanded into an energy formula featured by second, fourth, and sixth-order phenomenological anisotropy constants. After applying a temperature rescaling strategy, we show that the calculated Curie temperature, spin-reorientation phenomenon, $M_\text{s}(T)$, $\delta_w(T)$, and $K_i^\text{eff}(T)$ agree well with the experimental results. 
$A_\text{e}(T)$ is estimated through a general continuum description of the domain wall profile by mapping atomistic magnetic moments to the macroscopic magnetization. $A_\text{e}$ is found to decrease more slowly than $K_1^\text{eff}$ with increasing temperature, and approximately scale with normalized magnetization as $A_\text{e}(T) \sim m^{1.2}$. Especially, the possible domain wall configurations at temperatures below the spin-reorientation temperature and the associated  $\delta_w$ and $A_\text{e}$ are identified.
This work provokes a scale bridge between \textit{ab-initio} calculations and temperature-dependent micromagnetic simulations of Nd-Fe-B permanent magnets.
\end{abstract}

\maketitle

\section{Introduction}
Nd-Fe-B permanent magnets are critical for the key components of energy-related technologies, such as wind turbines and electro-mobility. They are also important in robotics, automatisation, sensors, actuators, and information technology \cite{gutfleisch2011magnetic,skokov2018heavy,hono2018prospect}. Since there is increasing demand in high-end technology that permanent magnets be used at finite or elevated temperatures, the temperature-dependent properties of Nd$_2$Fe$_{14}$B, the main phase of Nd-Fe-B magnets, are of great interest. For example, these magnets are exposed to elevated temperatures in many applications such as the motors inside hybrid vehicles where the operating temperature can approach 450 K.

Modelling and simulation play an important role in the design of permanent magnets for applications at elevated temperatures. Currently, first-principles calculations and micromagnetic simulations dominate the modelling of permanent magnets. The former helps to understand the magnetic properties on the electronic-level, as well as to predict intrinsic parameters (e.g. magnetic moment, crystal field parameter, etc.) at zero temperature \cite{liu2012partitioning,toga2015effects, suzuki2014effects,tatetsu2016first,yi2017multiscale,tsuchiura2014first}. However, first-principles calculations become very challenging at finite temperature. Micromagnetic model aims at simulating the domain structure on the nano/microscale level, and is very useful when studying the influence of microstructure (e.g. grain shape/size, grain boundary, intergranular phase, etc.) on the magnetization reversal process and the macroscopic properties of permanent magnets \cite{hrkac2010role,woodcock2012understanding,sepehri2014micromagnetic,fischbacher2018micromagnetics,fidler2000micromagnetic,hrkac2010role, yi2016micromagnetic,toson2016micromagnetic,helbig2017experimental,
zickler2017combined,erokhin2017optimization,fischbacher2018micromagnetics}. The thermal activation of nucleation at finite temperatures and its effect on the decay of coercive field in Nd-Fe-B magnets are addressed by micromagnetic simulations \cite{bance2015thermally,bance2015thermal}, but the temperature-dependent intrinsic properties have to be already known or determined beforehand. In addition, it is well known that the micromagnetic model is essentially a continuum approximation and assumes the magnetization to be a continuous function of position. This approximation holds when the considered length scales are large enough for the atomic structure to be ignored \cite{fidler2000micromagnetic,hirosawa2017perspectives}. However, when the region of interest is at the same scale as the exchange length, this approximation would fail. For example, in Nd-Fe-B magnets, the amorphous grain boundary is often found to be around 2 nm (close to the micromagnetic exchange length of Nd$_2$Fe$_{14}$B). The validity of micromagnetic representation of this 2-nm region with homogenized parameters remains as an issue. A scale bridge between these two methodologies for modeling Nd-Fe-B magnets is desired. Moreover, the evaluation of temperature-dependent macroscopic parameters for micromagnetic simulations is highly nontrivial. In this aspect, there are recent attempts to study temperature-dependent effective magnetic anisotropy, saturation magnetization, and reversal process in Nd$_2$Fe$_{14}$B by using atomistic spin model simulations \cite{evans2016atom,toga2016monte, nishino2017atomistic,tsuchiura2018bridging,miyashita2017perspectives}, based on which the concept of a multiscale model approach for the design of advanced permanent magnets is proposed \cite{westmoreland2018multiscale}. In general, an atomistic spin model is capable of calculating magnetic properties at different temperatures \cite{skubic2008method,evans2014atomistic,eriksson2017atomistic}, in which the temperature effects can be taken into account by either Langevin-like spin dynamics or Monte Carlo simulations. Its application to permanent magnets, or more especially rare-earth permanent magnets, is still at its early stage.
More efforts have to be made to either understand the gap between model simulations and experimental measurements or predict parameters over a broad range of temperatures, in order to establish the atomistic spin model as a readily available methodology for designing Nd-Fe-B magnets. In this work, following the similar framework in \cite{toga2016monte,nishino2017atomistic,toga2018anisotropy}, we not only calculate the Curie temperature and the temperature dependent magnetization, magnetocrystalline anisotropy and domain wall width, but also add some additional new knowledge into the community of Nd-Fe-B magnets in terms of atomistic spin model simulations and temperature dependent intrinsic parameters. For example, considering the different description of spin states in the classical and quantum manner, such as the different availability of spin states in the classical atomistic spin model simulations and experiments, we determine the temperature rescaling parameter for Nd$_2$Fe$_{14}$B and figure out the difference between simulation and experimental temperatures. In this way, the calculated magnetization \textit{vs} temperature curve shows a better agreement with the experimental one than that in \cite{toga2016monte}. In addition, except for the domain wall width at temperatures higher than the spin reorientation temperature, we also carefully examine various types of domain wall configurations and their width at temperatures lower than the spin reorientation temperature. Moreover, linking the simulation results and the micromagnetic theory, we determine the exchange stiffness for a wide range of temperatures and identify the scaling law.

Specifically, here we present an \textit{ab-initio} informed atomistic spin model for the theoretical calculation of the Curie temperature, spin-reorientation temperature, and magnetic properties of Nd$_2$Fe$_{14}$B, such as saturation magnetization $M_\text{s}(T)$, effective magnetic anisotropy constants $K_i(T)$ ($i=1,2,3$), domain wall width $\delta_w(T)$, and exchange stiffness constant $A_\text{e}(T)$ at temperatures both higher and lower than the spin reorientation temperature. The calculation results are coherent with the experimental results. Our work here provides effective parameters for micromagnetic simulations and will be useful for revealing the atomic-scale magnetic behavior in Nd-Fe-B magnets.

\section{Atomistic spin model for Nd$_2$Fe$_{14}$B}
\begin{figure}[!b]
\centering
\includegraphics[width=8.2cm]{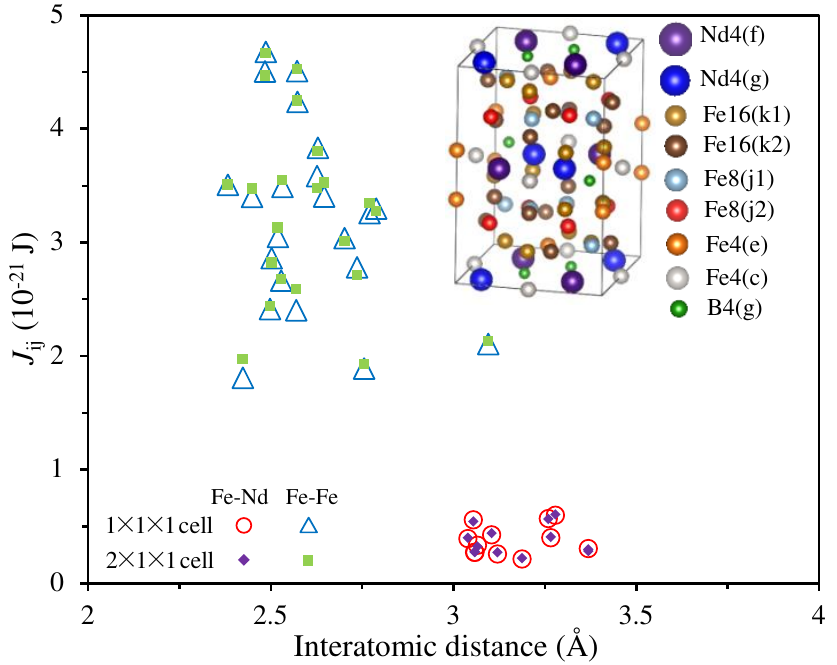}
\caption{Exchange parameters $J_{ij}$ as a function of interatomic distance, with the nearest neighbor considered. Inset: unit cell of Nd$_2$Fe$_{14}$B showing different kinds of crystallographically equivalent atoms. The results of  $2\times1\times1$ supercell are also presented to show the independence of $J_{ij}$ on the calculated cell size.}
\label{f1}
\end{figure}

For calculating the temperature-dependent magnetic properties, we use the atomistic spin model which treats each atom as a classic spin \cite{skubic2008method,evans2014atomistic,eriksson2017atomistic}. For Nd$_2$Fe$_{14}$B, the atomistic spin Hamiltanion can be written as
\begin{equation}
\begin{split}
 \mathcal{H} =  &
-\frac{1}{2}\sum_{i \neq j}^{i,j \in \text{Fe}}J_{ij}^\text{Fe-Fe}\mathbf{s}_{i} \cdot \mathbf{s}_{j} 
  -\frac{1}{2}\sum_{i \in \text{Fe}}^{j \in \text{Nd}} J_{ij}^\text{Fe-Nd}\mathbf{s}_{i} \cdot \mathbf{s}_{j}  \\
&-\sum_{i \in \text{Fe}} k_i^\text{Fe}(\mathbf{s}_i \cdot \mathbf{e}^z)^2
+\mathcal{H}_\text{Nd}^\text{cf}.
\end{split}
\label{eq1}
\end{equation}
It should be noted that in Eq. \ref{eq1} the energy terms from the external magnetic field and the dipole interaction between atomic spin moments are not included, since here we only focus on the calculation of intrinsic properties. $\mathbf{s}_{i}$ is a unit vector denoting the local spin moment direction. The first two terms in Eq. \ref{eq1} correspond to the Heisenberg exchange energy. They only contain the exchange interactions in Fe-Fe ($J_{ij}^\text{Fe-Fe}$) and Fe-Nd ($J_{ij}^\text{Fe-Nd}$) atomic pairs, owing to the fact that B sites are usually taken to be nonmagnetic and the interaction between Nd sites can be negligible \cite{toga2016monte, nishino2017atomistic,westmoreland2018multiscale}. The third term in Eq. \ref{eq1} represents the uniaxial magnetic anisotropy energy of Fe atoms, with $k_i^\text{Fe}$ as the anisotropy energy of per Fe atom and $\mathbf{e}^z$ the $z$-axis unit vector. The fourth term in Eq. \ref{eq1} denotes the crystal-field (CF) Hamiltonian of Nd ions, which is the main source of large magnetic anisotropy in Nd$_2$Fe$_{14}$B and can be approximated as \cite{yamada1988crystal,toga2016monte,nishino2017atomistic}
\begin{equation}
\mathcal{H}_\text{Nd}^\text{cf}=\sum_{i \in \text{Nd}} \sum_{n=2,4,6} \alpha_n \langle r^n \rangle_{4f,i} A_{n,i}^0 \hat{\Theta}_{n,i}^0,
\label{eq2}
\end{equation}
in which $\alpha_n$ is the Stevens factors, $\langle r^n \rangle_{4f,i}$ the 4$f$ radial expectation value of $r^n$ at the respective Nd site $i$, $A_{n,i}^0$ the CF parameters, and $\hat{\Theta}_{n,i}^0$ the Stevens operator equivalents.
For Nd$^{+3}$ ions, $\alpha_2=-6.428 \times 10^{-3}$, $\alpha_4=-2.911 \times 10^{-4}$, and $\alpha_6=-3.799 \times 10^{-5}$ \cite{elliott1953theory}. $\langle r^n \rangle$ values of Nd$^{+3}$ ions can be calculated as $\langle r^2 \rangle=1.001 a_0^2$, $\langle r^4 \rangle=2.401 a_0^4$, and $\langle r^6 \rangle=12.396 a_0^6$ in which $a_0$ is the Bohr radius \cite{freeman1962theoretical}.
The Stevens operator equivalents are expressed as \cite{elliott1953theory}
\begin{equation}
\begin{split}
\hat{\Theta}_2^0 = & 3J_z^2 - \mathcal{J} \\
\hat{\Theta}_4^0 = & 35J_z^4 - 30\mathcal{J}J_z^2 + 25J_z^2 - 6\mathcal{J} + 3\mathcal{J}^2 \\
\hat{\Theta}_6^0 = & 231J_z^6 - 315\mathcal{J}J_z^4 + 735J_z^4 + 105\mathcal{J}^2J_z^2 \\ 
& - 525\mathcal{J}J_z^2 + 294J_z^2 - 5\mathcal{J}^3 + 40\mathcal{J}^2 - 60\mathcal{J} .
\end{split}
\label{eq3}
\end{equation}
$J_z=J(\mathbf{s} \cdot \mathbf{e}^z)$ denotes the $z$-component of the total angular momentum $J$ which is 9/2 for Nd ions \cite{elliott1953theory}. $\mathcal{J}=J^2$ instead of $\mathcal{J}=J(J+1)$ is used in the classical manner \cite{toga2016monte}. The reliable first-principles calculation of high-order CF parameters in Nd$_2$Fe$_{14}$B is still challenging. Here we take the $A_n^0$ values which are determined from the experiment results \cite{yamada1988crystal}, i.e. $A_2^0=295$ K/$a_0^2$, $A_4^0=-12.3$ K/$a_0^4$, and $A_6^0=-1.84$ K/$a_0^6$. We approximately set all Nd ions with the same CF parameters. In this way, combining Eq. \ref{eq2}, Eq. \ref{eq3}, and $J_{i,z}=J(\mathbf{s}_i \cdot \mathbf{e}^z)$ yields the CF energy
\begin{equation}
\begin{split}
& \mathcal{H}_\text{Nd}^\text{cf}=  \\
& -\sum_{i \in \text{Nd}} \left[ k_{i,1}^\text{Nd}(\mathbf{s}_i \cdot \mathbf{e}^z)^2
+ k_{i,2}^\text{Nd}(\mathbf{s}_i \cdot \mathbf{e}^z)^4 + k_{i,3}^\text{Nd}(\mathbf{s}_i \cdot \mathbf{e}^z)^6
 \right],
\label{eq4}
\end{split}
\end{equation}
in which the parameters $k_{i,1}^\text{Nd}$, $k_{i,2}^\text{Nd}$, and $k_{i,3}^\text{Nd}$ are listed in Table \ref{tab1}. The constant term in $\mathcal{H}_\text{Nd}^\text{cf}$ is not important and thus not presented in Eq. \ref{eq4}.
The magnetocrystalline anisotropy energy of the Fe sublattice and the magnetic moments of each atom, as listed in Table \ref{tab1}, are taken from the previous first-principles calculations \cite{miura2014magnetocrystalline,toga2016monte}.
The exchange parameters $J_{ij}^\text{Fe-Fe}$ and $J_{ij}^\text{Fe-Nd}$ in Eq. \ref{eq1} are evaluated in the relaxed unit cell (lattice parameters are kept constant as $a=b=8.76$ \AA, $c=12.13$ \AA \, and the thermal expansion is not considered) by using OpenMX \cite{liechtenstein1987local, han2004electronic,yoon2018reliability,kim2018calculating}.
The calculation of Heisenberg exchange parameters $J_{ij}$ between two different atomic sites $i$ and $j$ is implemented in OpenMX by using the magnetic-force theorem (follow the original formalism by Liechtenstein et al. \cite{liechtenstein1987local}) and its extension to the nonorthogonal LCPAO (linear combination of pseudoatomic orbitals) method \cite{han2004electronic}. In detail, $J_{ij}$ is estimated as a response to small spin tiltings (as a perturbation) from the given converged solution, as shown the detailed formulation in \cite{liechtenstein1987local, han2004electronic,yoon2018reliability}. More application examples of OpenMX in calculating Heisenberg exchange parameters are reported by the OpenMX's developers in the literature \cite{han2004electronic,yoon2018reliability,kim2018calculating,jang2018charge}.
In fact, the unit cell here is already very large and thus the lattice translation vectors have negligible influence on the calculated $J_{ij}$. Indeed, our additional calculations of the $2\times1\times1$ and $2\times2\times1$ supercells show that the influence of the adopted cell size on the calculated $J_{ij}$ can be ignored, as shown in Fig. \ref{f1}. Therefore, the calculated $J_{ij}$ here can be used in the Heisenberg spin model and the Monte Carlo simulations.
An open-core pseudopotential for Nd is used, with the 4$f$ electrons put in the core and not treated as valence electrons. For the many local-orbital-based methods in OpenMX, the basis set of each atom should be chosen. We use a notation of $sN_spN_pdN_dfN_f$ to represent the basis-set choice for a given atom. For example, $s1p2d3$ denotes that one $s$, two $p$, and three $d$ orbitals are taken as a basis set. According to the previous work  \cite{tatetsu2016first}, the basis sets for Nd, Fe, and B atoms are chosen as $s2p2d2$, $s2p2d2$, and $s2p2$, with cutoff radii of 8.0, 6.0, and 7.0 a.u., respectively. We use a $5\times5\times4$ $k$-point mesh, and a 500-Ry cutoff energy. The convergence criteria for the selfconsistent calculation is 10$^{-6}$ Hartree. The calculated exchange parameters are further calibrated (interactions of Fe--Fe and Fe--Nd are rescaled by 2 and 0.9, respectively) by checking the results from the atomistic spin model simulation of Nd$_2$Fe$_{14}$B, and are shown in Fig. \ref{f1}. It can be found that the total magnetic moment of Nd ions is ferromagnetically coupled to Fe moments, and the exchange of Fe-Fe pairs is 3$-$10 times stronger as that of Fe-Nd pairs.
Previous studies have shown that the cutoff radius (within which exchange parameters are calculated) affect the magnetization at higher temperatures \cite{toga2016monte}. In order to reduce the computational cost, as a simplification, here we only calculate exchange parameters within the nearest-neighbor approximation. The effect from longer-range exchange interactions is not included. For Nd$_2$Fe$_{14}$B system, the nearest-neighbor exchange interactions dominate while the longer-range ones are less important. 
In the following we will show that the calculated macroscopic properties from this simplification are in line with the previous work \cite{toga2016monte} and the experimental report \cite{durst1986determination,hirosawa1986magnetization,givord1993magnetic}, without significant disparity.
It should be noted that the micromagnetic exchange length is evaluated from the micromagnetic model in the framework of continuum picture without information from the atomistic spin at each atomic site. The micromagnetic exchange length governs the width of the transition between magnetic domains. In contrast, the exchange parameters describe the interaction between each pair of atomistic spins at specific atomic sites. They are in the framework of  discrete picture in the atomistic spin scale. Thus, the micromagnetic exchange is not a direct indicator for the cutoff radius of the exchange interaction in the atomistic spin model.

\begin{table}[!b]
\caption{Magnetic moments and atomic-site resolved magnetic anistropy energy of each crystallographically equivalent atom.}
\newcommand{\tabincell}[2]{\begin{tabular}{@{}#1@{}}#2\end{tabular}}
\begin{center}
 \label{tab1} 
\begin{tabular}{cccc}
 \hline
\quad Atom \quad & \quad \tabincell{c}{$\mu_i$ \\ ($\mu_\text{B}$) } \quad & \quad \tabincell{c}{$k_i$ \\ ($\times 10^{-21}$ J)} \\
 \hline
 &  & $k_i^\text{Nd}$ & \\
\tabincell{c}{Nd($4g$) \\ Nd($4f$)} & \tabincell{c}{2.86 \\ 2.871 } & \tabincell{c}{$k_{i,1}^\text{Nd}=-4.935$ \\  $k_{i,2}^\text{Nd}=25.98$ \\ $k_{i,3}^\text{Nd}=-22.94$}  \\ 
 &  & $k_i^\text{Fe}$ & \\ 
Fe($4c$) & 2.531 & $-0.342$ & \\ 
Fe($4e$) & 1.874 & $-0.0048$ & \\ 
Fe($8j_2$) & 2.629 & $0.093$ &\\ 
Fe($8j_1$) & 2.298 & $0.171$ &\\ 
Fe($16k_2$) & 2.206 & $0.0608$ & \\ 
Fe($16k_1$) & 2.063 & $0.0880$ &\\ 
 \hline
\end{tabular}
\end{center}
\end{table}

\begin{figure*}[!t]
\centering
\includegraphics[width=14.6cm]{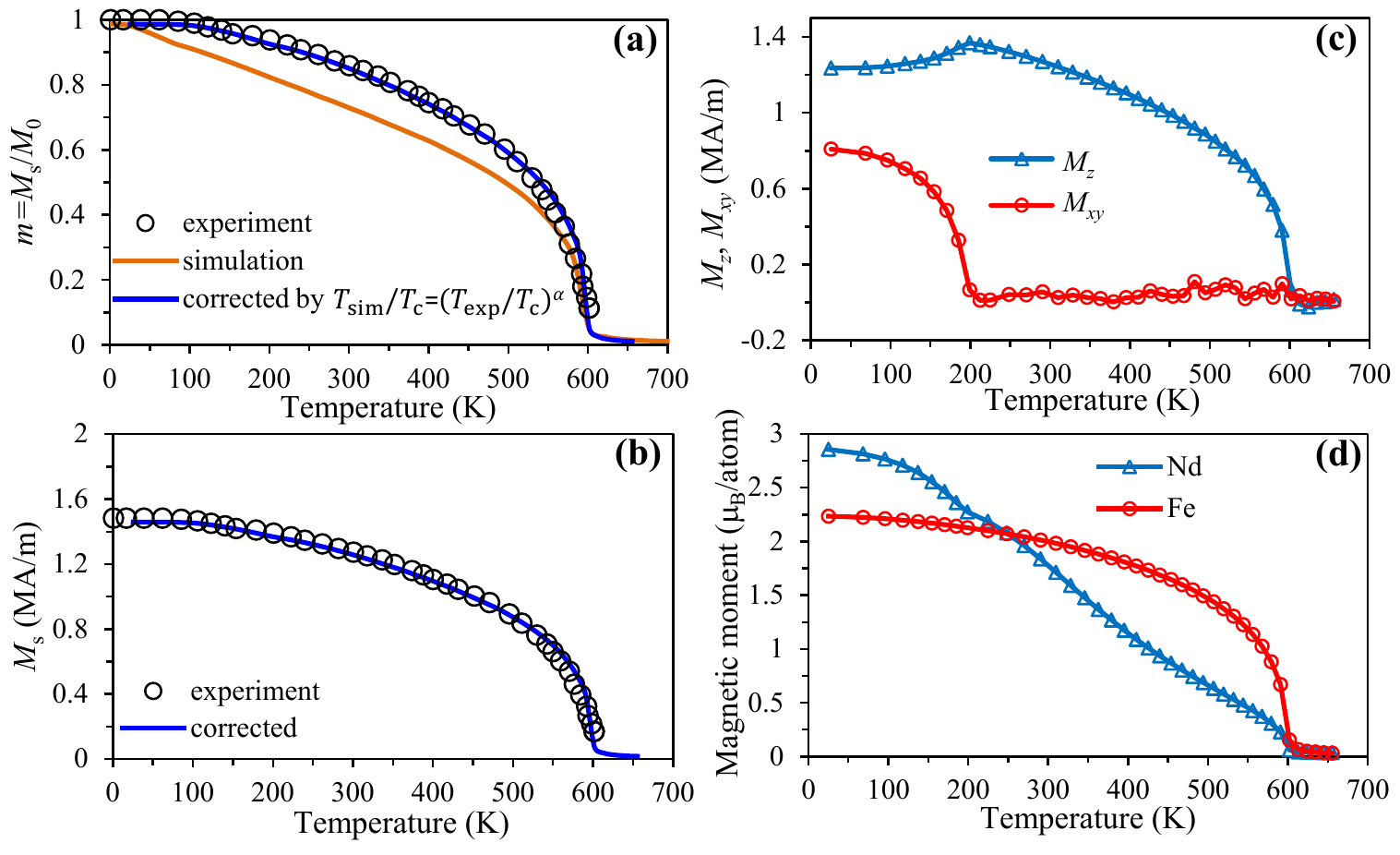}
\caption{Temperature dependence of (a) (b) the magnetization amplitude, (c) the magnetization components $M_z$ and $M_{xy}$, and (d) the magnetic moment per atom in Nd and Fe sublattices. The corrected curves are plotted by $\alpha=1.802$. The experimental results are taken from \cite{hirosawa1986magnetization}.}
\label{f2}
\end{figure*}

After parameterization, the atomistic spin model in Eq. \ref{eq1} is implemented in VAMPIRE \cite{evans2014atomistic}. For calculating the Curie temperature and temperature-dependent magnetization, the Monte Carlo Metropolis method is adopted, using a sample with $10\times10\times10$ unit cells and periodic boundary conditions in all three directions. After performing 10,000 Monte Carlo steps at each temperature, the equilibrium properties of the system are calculated by averaging the magnetic moments over a further 10,000 steps. It should be noted that by performing calculations at different steps, we find the results remain the same after Monte Carlo steps exceed 10,000. For the calculation of effective magnetic anisotropy constants at different temperatures, we use the constrained Monte-Carlo method \cite{evans2014atomistic,asselin2010constrained}. We constrain the direction of the global magnetization at a fixed polar angle ($\theta$) while allow the individual spins to vary. In this way, we can calculate the restoring torque acting on the magnetization as a function of $\theta$, from which the effective magnetic anisotropy constants can be obtained by fitting. When calculating domain wall width, we apply the spin dynamics approach and the Heun integration scheme. A sharp Bloch-like domain wall (wall plane perpendicular to $x$ axis) in the middle of the sample with $N_x \times N_y \times N_z$ unit cells is set as the initial condition.
The system with the demagnetizing field included further relaxes from this initial condition by 100,000 steps with a time step of 1 fs. The final domain configuration is determined by averaging the magnetic moment distribution of 100 states at 90.1, 90.2, 90.3, $\dots$, 100 ps.

\section{Results and discussion}
\subsection{Curie temperature and saturation magnetization}
The calculated temperature-dependent magnetization curve for Nd$_2$Fe$_{14}$B is shown in Fig. \ref{f2}(a). For a classical spin model, the simulated magnetization can be related to temperature through the function \cite{evans2014atomistic}
\begin{equation}
m=M_\text{s}(T)/M_0 = (1-T/T_\text{c})^\beta ,
\label{eq5}
\end{equation}
in which $M_\text{s}(T)$ is the temperature dependent saturation magnetization, $M_0$ denotes the saturation magnetization at zero K, $T_\text{c}$ is the Curie temperature, and $\beta$ is an exponent. Direct fitting the simulation data by Eq. \ref{eq5} gives $T_\text{c}=602$ K and $\beta=0.418$. The calculated $T_\text{c}$ matches well with the experimental data \cite{hirosawa1986magnetization}.

However, it can be found from Fig. \ref{f2}(a) that only the simulation results around the Curie temperature agree with the experimental measurement.
This disparity could be related to the following two aspects. Firstly, the exchange parameters could vary when temperature changes, as the case for Fe shown in \cite{ruban2004atomic}. At high temperatures, there may exist disordered local moment (DLM) state \cite{staunton1986static} and thus different exchange parameters and magnetization. However, the calculation of temperature dependent exchange parameters by first-principles methods is still challenging for the complicated Nd$_2$Fe$_{14}$B. Nevertheless, using the constant exchange parameters, the Curie temperature of Nd$_2$Fe$_{14}$B is well predicted in Fig. \ref{f2}(a). Apart from the possible reason related to temperature or DLM-state dependent exchange parameters, we think the distinction between the quantum model and the classical model should also contribute to the deviation in Fig. \ref{f2}(a), as thoroughly discussed in \cite{evans2015quantitative}. Atomistic spin model is a classical model which considers localized classical atomistic spins with unrestricted and continuous values. In contrast, the experimental measurement spontaneously includes the manifestation of a quantum system which only allows particular eigenvalues. It indicates more available states in the classical model than in experiments. The macroscopic magnetization obtained at simulation temperature $T_\text{sim}$ should be achieved at higher temperature $T_\text{exp}$ in experiments. For this reason, there should be a mapping between $T_\text{sim}$ and $T_\text{exp}$. Here we adopt the temperature rescaling method, as proposed in the previous work \cite{evans2015quantitative}, to determine this mapping.
The (internal) simulation temperature $T_\text{sim}$ is rescaled so that the equilibrium magnetization at the input experimental (external) temperature $T_\text{exp}$ agrees with the experimental result, i.e.
\begin{equation}
T_\text{sim}/T_\text{c}=\left(T_\text{exp}/T_\text{c} \right)^\alpha,
\label{eq6}
\end{equation}
in which $\alpha$ is the rescaling parameter which can be fitted. 
The physical interpretation of the rescaling is that at low temperatures the allowed spin fluctuations in the classical limit are overestimated, and so this corresponds to a higher effective temperature than given in the simulation (i.e. $T_\text{exp}>T_\text{sim}$) \cite{evans2015quantitative}. 
The physical origin of $\alpha$ may be relate to the different availability of spin states in the classical atomistic spin model simulation and the experiment. However, it would be interesting to apply detailed first-principles calculations to delineate the origin.
For detailed discussion on the temperature rescaling, the readers are referred to \cite{evans2015quantitative}. Applying the temperature rescaling Eq. \ref{eq6} to the simulation data and directly comparing the rescaled data with the experimental data, we fit the parameter $\alpha$ as 1.802. After these operations, we can see in Fig. \ref{f2}(a) that the corrected simulation data show excellent agreement with the experimental one, and both can be described by the Curie--Bloch equation
\begin{equation}
m=M_\text{s}(T)/M_0=\left[ 1-(T/T_\text{c})^\alpha \right]^\beta
\label{eq7}
\end{equation}
with the fitted parameters $\alpha=1.802$ and $\beta=0.418$.

Calculating the total magnetic moments per volume, we then obtain the temperature dependent saturation magnetization $M_\text{s}(T)$ from the corrected simulation data. $M_\text{s}(T)$ agrees well with the experimental data \cite{hirosawa1986magnetization}, as shown in Fig. \ref{f2}(b). The spin reorientation phenomenon can also be captured by atomistic spin simulations, as shown in Fig. \ref{f2}(c). The simulated $M_z$ in Fig. \ref{f2}(c) firstly increases and then decreases with the increasing temperature. By comparing $M_z$ in Fig. \ref{f2}(c) to $M_\text{s}$ in Fig. \ref{f2}(b), it can be estimated that the tilting angle of the magnetization direction away from the $z$-axis is around 32$^\circ$ at $T=25$ K. The simulated spin reorientation temperature is around 180 K, higher than the experimental value around 150 K. This deviation may be related to the low quality of the temperature rescaling at low temperature. Nevertheless, the results on spin reorientation are in line with the experimental observations \cite{yamada1988crystal, hirosawa1986magnetization,givord1993magnetic}.
Meanwhile, it can be seen from Fig. \ref{f2}(d) that as temperature increases, the magnetization of Nd sublattice decreases faster than that of Fe sublattice. This is due to the strong exchange coupling in Fe sublattice and indicates that Fe sublattice is responsible for the magnetic order.

\begin{figure}[!t]
\centering
\includegraphics[width=8.2cm]{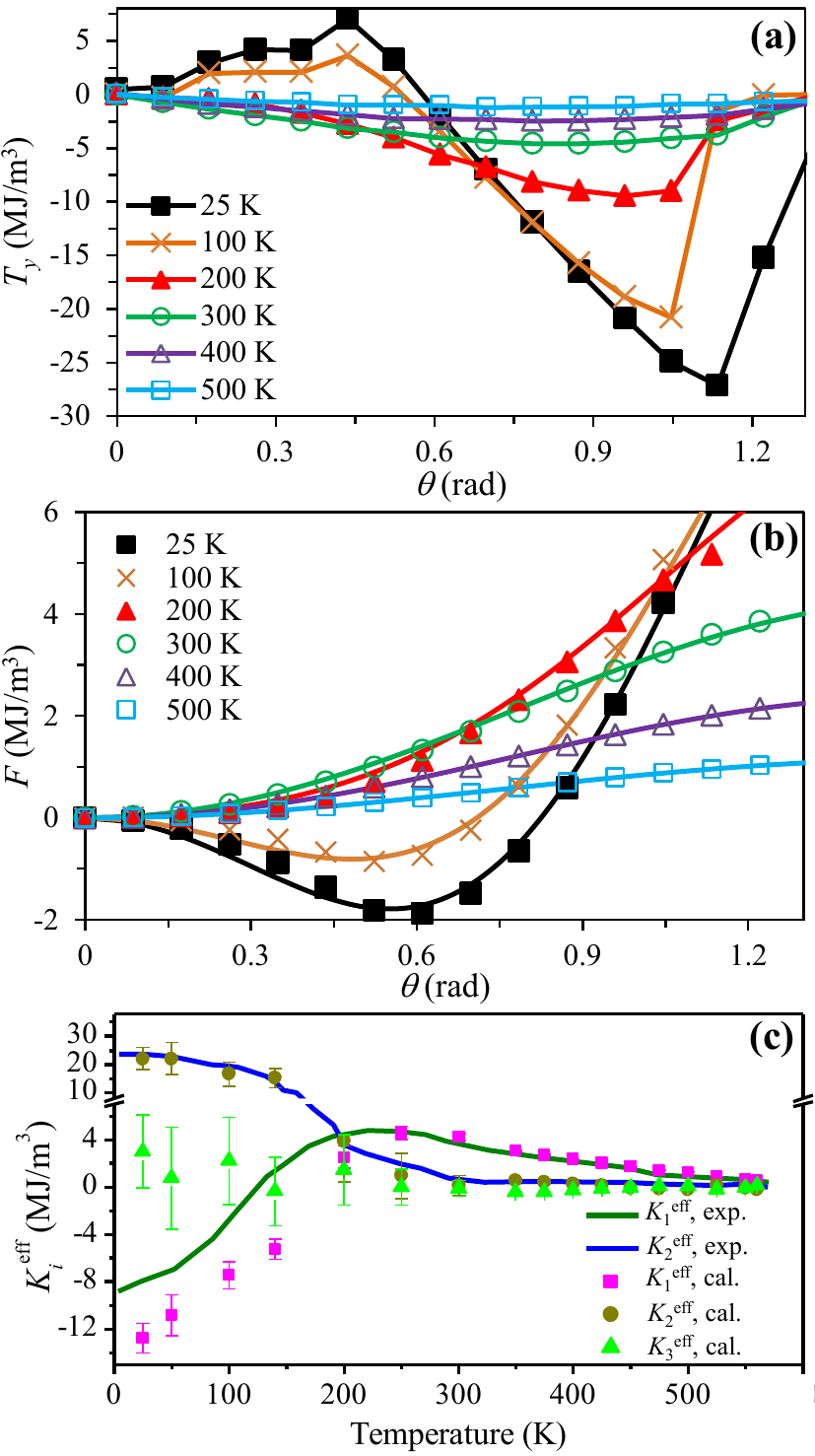}
\caption{(a) Internal torque density $T_y(\theta)$ and (b) free energy density $F(\theta)$ at different temperatures. (c) Temperature dependent experimental and calculated effective magnetic anisotropy constants $K_i^\text{eff}$ ($i=1,~2,~3$).}
\label{f3}
\end{figure}

\subsection{Effective magnetic anisotropy}
In order to determine the effective magnetic anisotropy constants, we have to calculate the system energy when the global magnetization is aligned along different directions. This can be done through the calculation of torque. In the constrained Monte Carlo scheme, we fix the azimuthal angle at zero degree and gradually change polar angle from 0 to 90 degree, i.e., the global magnetization is rotated in the $z$-$x$ plane and only the torque component $T_y$ is nonzero. The total internal torque $T_y$ is calculated from the thermodynamic average and transferred into the energy per volume, as shown in Fig. \ref{f3}(a). It can be seen that at low temperature (e.g. 25 and 100 K) $T_y$ is positive when $\theta$ is close to the the $z/[001]$ axis, indicating a spontaneous deviation of the global magnetization from the $z/[001]$ axis. This result is in line with the easy-cone type of anisotropy and the spin tilting away from $z/[001]$ axis (Fig. \ref{f2}(c)) at low temperature. At high temperature, $T_y$ is always negative and thus there is a revert torque for driving the global magnetization towards the $z/[001]$ axis, implying an easy-axis type of anisotropy.

After obtaining the temperature dependent $T_y$, the free energy ($F$) of the magnetic system can be related to the work done by the torque acting on the whole system, i.e.
\begin{equation}
F(\theta, T)=-\int_0^\theta T_y(\Theta,T) \text{d}\Theta.
\label{eq8}
\end{equation}
Integrating the data in Fig. \ref{f3}(a) through Eq. \ref{eq8} gives the free-energy curves in Fig. \ref{f3}(b). It can be seen that at 25 K, $F$ shows a local minimum at $\theta \approx 32^\circ$, reflecting the spin tilting away from $z/[001]$ axis. 
The effective magnetic anisotropy constants can be determined through the fitting of $F$ curves by the phenomenological six-order formula 
\begin{equation}
F(\theta, T)= K_1^\text{eff}(T)\sin^2\theta +  K_2^\text{eff}(T)\sin^4\theta +  K_3^\text{eff}(T)\sin^6\theta,
\label{eq9}
\end{equation}
in which $K_1^\text{eff}$, $K_2^\text{eff}$, and $K_3^\text{eff}$ are the macroscopically effective second-, fourth-, and sixth-order anisotropy constants, respectively. The fitting results are presented in Fig. \ref{f3}(c) and compared to the experimental measurement \cite{durst1986determination}. We can see that below 150 K, $K_1^\text{eff}$ is negative and both $K_2^\text{eff}$ and $K_3^\text{eff}$ play a critical role, agreeing with the cone-type anistropy of Nd$_2$Fe$_{14}$B at low temperature. After 250 K, $K_1^\text{eff}$ dominates and $K_2^\text{eff}$ and $K_3^\text{eff}$ are relatively small.
At 300 K, our calculated results are: $K_1^\text{eff}=4.26$ MJ/m$^3$, $K_2^\text{eff}=0.15$ MJ/m$^3$, and $K_3^\text{eff}=-0.10$ MJ/m$^3$. At higher temperature, $K_2^\text{eff}$ and $K_3^\text{eff}$ almost vanish. The calculated temperature dependence of $K_i^\text{eff}$ in Fig. \ref{f3}(c) agrees reasonably with the previous experimental measurement \cite{durst1986determination,yamada1986magnetocrystalline, hirosawa1986magnetization} and theoretical calculations \cite{toga2016monte,sasaki2015theoretical}.

\begin{figure*}[!t]
\centering
\includegraphics[width=15.6cm]{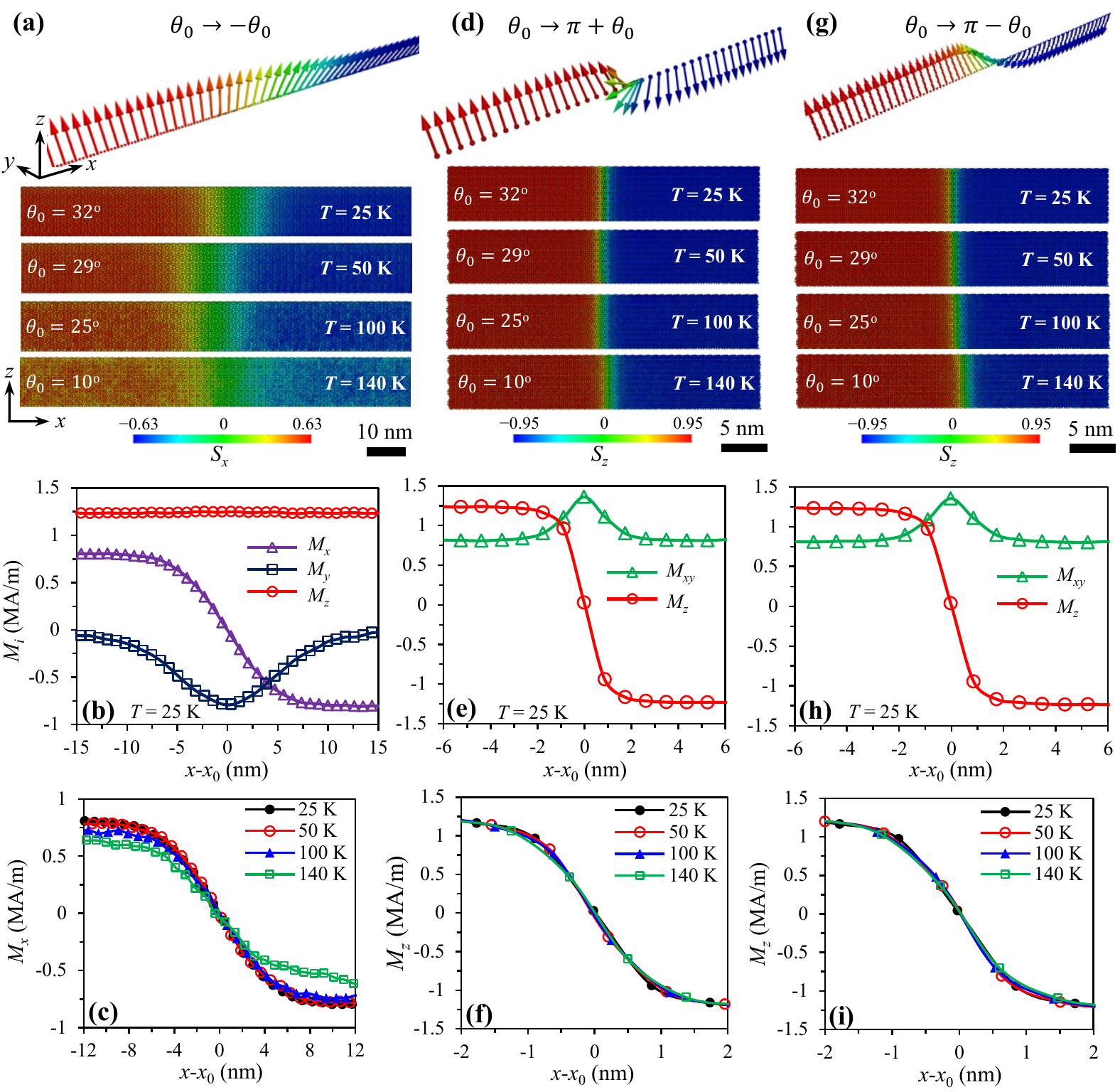}
\caption{(a) (d) (g) Three types of possible low-temperature (easy axis tilted from $z$-axis with angle $\theta_0$) domain wall configuration displayed by the distribution of atomistic magnetic moments. The distribution of macroscopic magnetization components along $x$ axis in the case of (b) (c) domain wall $\theta_0 \to -\theta_0$, (e) (f) domain wall $\theta_0 \to \pi+\theta_0$, and (h) (i) domain wall $\theta_0 \to \pi-\theta_0$.}
\label{f4}
\end{figure*}

\begin{figure*}[!t]
\centering
\includegraphics[width=15.6cm]{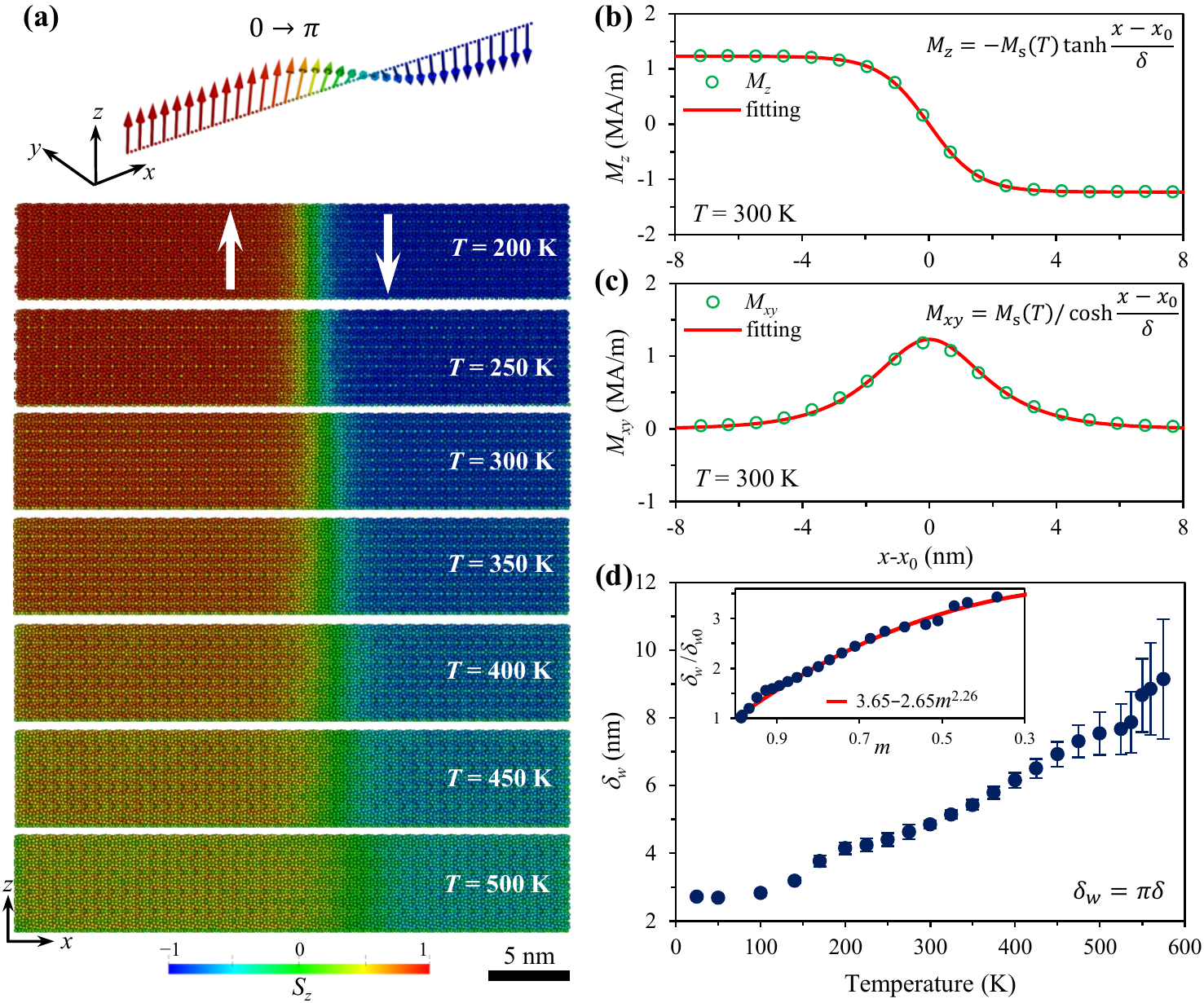}
\caption{(a) High-temperature (easy axis along $z$-axis) domain wall configuration displayed by the distribution of atomistic magnetic moments. Macroscopic (b) $M_z$ and (c) $M_{xy}$ distribution along $x$ axis at $T=300$ K. (d) Domain wall width $\delta_w$ at different temperatures. Inset in (d): $\delta_w$ scaling with magnetization as a function of $m^{2.92}$. $\delta_{w0}$ is the wall width at zero temperature.}
\label{f5}
\end{figure*}

\subsection{Domain wall}
Due to the different anisotropy types at low temperature (cone-type anisotropy) and high temperature (easy-axis anisotropy) in Nd$_2$Fe$_{14}$B, the domain wall will also be distinct. At temperatures lower than the spin reorientation temperature, a number of possible variants of domain-wall types have been observed due to the cone-type anisotropy \cite{pas1997temperature, pas1999magnetic}. 
For hard materials (Nd-Fe-B permanent magnets here) with dominant magnetocrystalline anisotropy, the typical domain wall profile is of the Bloch type, i.e. the magnetization is parallel to the easy axis ($z$ or $c$ axis for Nd$_2$Fe$_{14}$B) in the two domains separated by a domain wall perpendicular to $x$ ($a$) axis. Hence, here we study the Bloch-like domain walls, with the wall plane perpendicular to $x$ axis, as shown in Figs. \ref{f4} and \ref{f5}.
We consider three types of Bloch-like domain walls at low temperatures in Fig. \ref{f4}. More complicated domain walls with the wall plane perpendicular to different crystallographic axes will be investigated in our next work. The three wall modes are described as the polar angle changing from $\theta_0$ to $-\theta_0$ in Fig. \ref{f4}(a), $\theta_0$ to $\pi+\theta_0$ in Fig. \ref{f4}(d), and $\theta_0$ to $\pi-\theta_0$ in Fig. \ref{f4}(g), with the angle through the wall as $2\theta_0$, $\pi$, and $\pi-2\theta_0$, respectively.
At temperatures higher than the spin reorientation temperature, the 180$^\circ$ Bloch-like domain wall with the polar angle changing from 0 to $\pi$ is considered, as shown in Fig. \ref{f5}(a).

For calculating the domain wall, we set the magnetic moment direction in the $y-z$ plane with a polar angle as $\theta_0$ (i.e. tilting angle) and $-\theta_0$ (or $\pi \pm \theta_0)$ in the upward and downward domain, respectively. We then relax the system to attain the distribution of magnetic moments around the domain wall, as shown in Fig. \ref{f4}(a)(d)(g) and Fig. \ref{f5}(a). It can be seen that at low temperature (e.g. below 200 K) the magnetic moments are uniformly distributed within the domain, and a clear transition of magnetic moment distribution from the domain wall to the domain is visually observable. In contrast, at higher temperatures (e.g. above 400 K), the effect of thermal fluctuations is stronger, so that there are some randomly distributed magnetic moments even in the domain and no obvious transition between the domain wall and domain can be intuitively identified.

In order to determine the domain wall width, we turn to the continuum description of domain wall or diffusive interface. For mapping the atomistic magnetic moments to the continuum magnetization, we divide the simulation sample with $N_x \times N_y \times N_z=40 \times 5 \times 5$ unit cells into $N_x$ parts along $x$ axis. For the case of $2\theta_0$ domain wall in Fig. \ref{f4}(a), the wall is very wide and thus a simulation sample with $N_x \times N_y \times N_z=120 \times 5 \times 5$ unit cells is used. Each part (with an index of $l_x$, $1 \le l_x \le N_x$) represents $1 \times 5 \times 5$ unit cells, with its $x$ coordinate set in its center. The magnetization of each part is calculated by dividing its total magnetic moments by its volume. In this way, we attain the magnetization components $M_i(x)$ and $M_{jk}(x)$ for each part $l_x$ from the atomistic results in Figs. \ref{f4} and \ref{f5}, i.e.
\begin{equation}
M_i(x)=\sum_{I \in l_x} \frac{\mu_I s_I^i}{V_{l_x}} \label{eq10}
\end{equation}
and
\begin{equation}
M_{jk}(x)=\sum_{I \in l_x} \frac{\mu_I \sqrt{(s_I^j)^2+(s_I^k)^2}}{V_{l_x}} \label{eq11}
\end{equation}
at $x=0.5+(l_x -1)a$, in which $\mu_I$ is the magnetic moment of atom $I$ in the part $l_x$, $s_I^i$ ($i=x,y,z$) the spin direction components of atom $I$, $V_{l_x}$ the volume of part $l_x$, and $a=8.76$ \AA ~the in-plane lattice parameter. Following the mapping in Eq. \ref{eq10}, we obtain the scattered data to describe the domain wall configuration, as shown in Figs. \ref{f4} and \ref{f5}.
In the continuum model, the domain wall or diffusive interface can be described by the hyperbolic functions \cite{sun2007sharp,hubert2008magnetic,manfred2003micromagnetism} through
\begin{equation}
M_i(x)=-M_\text{s}(T)\tanh \frac{x-x_0}{\delta} \label{eq12}
\end{equation}
or
\begin{equation}
M_{jk}(x)=M_\text{s}(T)/\cosh \frac{x-x_0}{\delta}, \label{eq13}
\end{equation}
in which $x_0$ is for shifting the domain wall to the center and $\delta$ is the parameter related to domain wall width $\delta_w$ by $\delta_w=\pi \delta$.

In Fig. \ref{f4}, we present the domain wall profile at temperature lower than the spin reorientation temperature. For the $2\theta_0$ domain wall in Fig. \ref{f4}(a), the domain wall width is quiet large. It can be found from Fig. \ref{f4}(b) and (c) that $M_z$ does not change along $x$ axis, whereas $M_x$ can be described by Eq. \ref{eq12}. So this wall does not satisfy the condition of constant normal component of the magnetization along the wall axis, i.e. not a Bloch-like wall. Moreover, the uniform $M_z$ indicates constant magnetic anisotropy energy according to Eq. \ref{eq9} and thus the domain wall cannot exist; because the formation of domain wall is a result of the competition between variable exchange energy and magnetic anisotropy energy. One possible explanation for the wide domain wall in Fig. \ref{f4}(a) is that, the azimuthal angle also takes effects in the magnetic anisotropy energy and could contribute to the domain wall formation. The role of azimuthal angle in determining the easy direction of Nd$_2$Fe$_{14}$B at low temperatures has also been addressed before \cite{pas1999magnetic}. However, in Eq. \ref{eq9} we neglect the azimuth-angle dependence, which has to be taken into account in the following work. Here we focus on the Bloch-like wall and will not put emphasis on the wide domain wall in Fig. \ref{f4}(a) as well as its width.
In contrast, the $\pi$ and $\pi-2\theta_0$ domain walls are Bloch-like and narrow, and $M_z$ can be well described by Eq. \ref{eq12}, as shown in Fig. \ref{f4}(e), (f), (h) and (i). The domain wall becomes slightly wider as the temperature increases from 25 K to 140 K. In addition, the wall profiles in $\pi$ and $\pi-2\theta_0$ domain walls are almost the same at a specific temperature. In the following, we will take the wall profile in $\pi-2\theta_0$ domain wall to calculate the domain wall width and exchange stiffness at low temperatures.

At temperatures higher than the spin reorientation temperature, 180 degree Bloch-like domain walls clearly form, as shown in Fig. \ref{f5}(a). Fitting the scattered data associated with the domain wall configuration by Eq. \ref{eq12} or \ref{eq13} can give $\delta$ and thus the domain wall width. Typical fitting results at 300 K are presented in Fig. \ref{f5}(b) and (c), with $\delta=1.55$ nm and $\delta_w=4.87$ nm. It should be noted that at 300 K, the exchange stiffness $A_\text{e}$ is often taken as 6.6--12 pJ/m \cite{ono2014observation,durst1986determination,manfred2003micromagnetism} and $K_1^\text{eff}$ as 4.2--4.5 MJ/m$^3$ \cite{durst1986determination, manfred2003micromagnetism} in the literature, corresponding to an estimated $\delta_w=\pi \sqrt{A_\text{e}/K_1^\text{eff}}$ as 3.63--5.31 nm. Our calculated $\delta_w$ at 300 K falls well in the range of $\delta_w$ estimated from the literature. The measured $\delta_w$ by electron microscopy is more widely distributed, ranging from 1 to 10 nm \cite{zhu1998magnetic, lloyd2002measurement,beleggia2007quantitative}.
The calculated domain wall width at different temperatures are summarized in Fig. \ref{f5}(d). It can be found that domain wall becomes wider as the temperature increases, from $\delta_w=2.72$ nm at 25 K to $\delta_w=8.67$ nm at 550 K. The large standard deviation of $\delta_w$ at higher temperature is attributed to the stronger thermal fluctuations. These results are also consistent with the previous simulation results \cite{nishino2017atomistic}.
In addition, the dimensionless wall width $\delta_w / \delta_{w0} $ ($\delta_{w0}$: wall width at 0 K) can be fitted as a function of the power of dimensionless magnetization, i.e. $\delta_w / \delta_{w0} $ linearly varies with $m^{2.26}$, as shown in the inset of Fig. \ref{f5}(d). This is different from the low-temperature power-law scaling behavior of $m^{-0.59}$ as found in cobalt \cite{moreno2016temperature}, possibly due to the complicated and intrinsically different crystal structure of Nd$_2$Fe$_{14}$B.

\begin{figure}[!b]
\centering
\includegraphics[width=8.4cm]{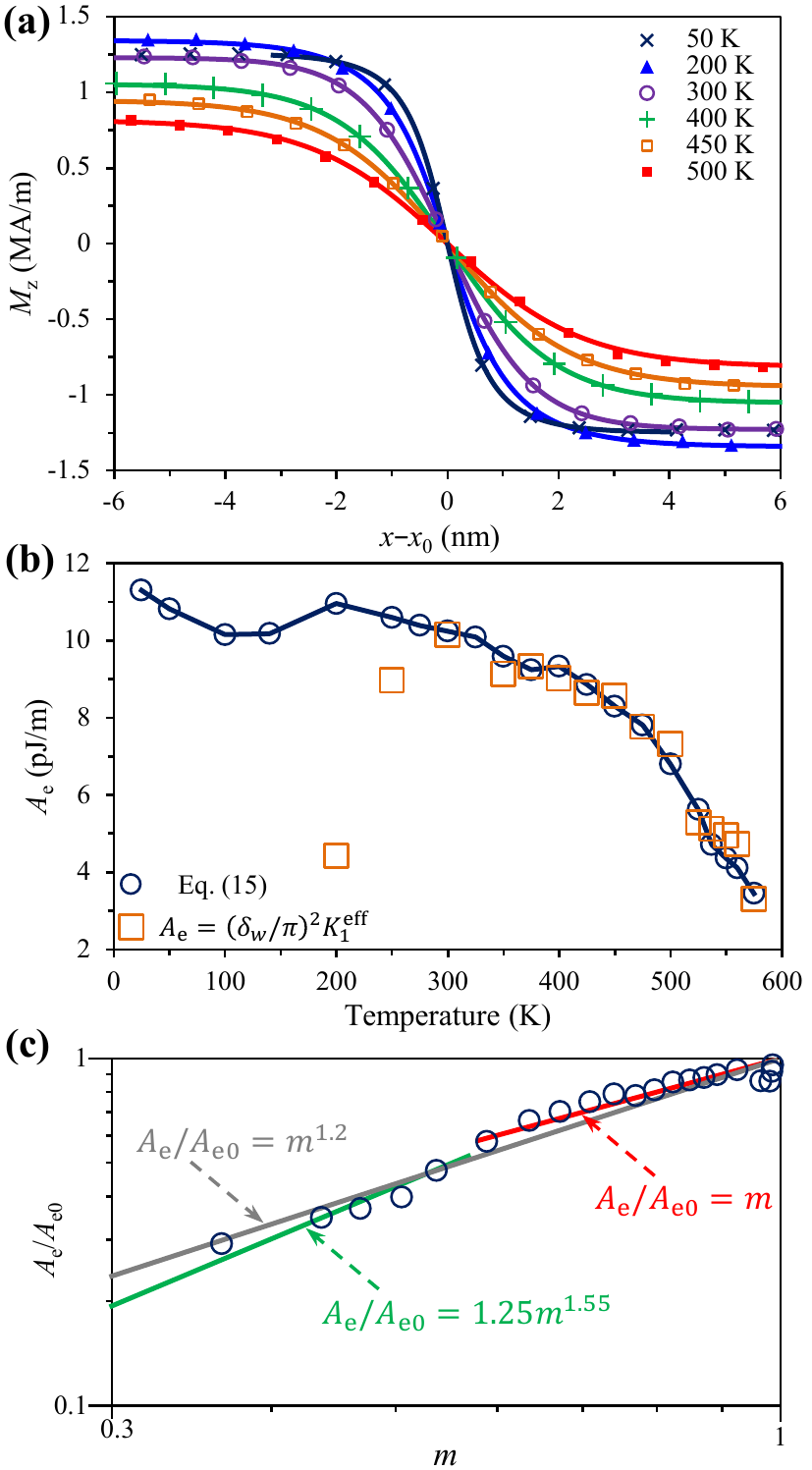}
\caption{(a) $M_z$ distribution along $x$ axis at different temperatures. (b) Calculated temperature-dependent exchange stiffness. (c) Scaling behavior of the exchange stiffness, with the solid lines showing the scaling law with normalized magnetization. $A_\text{e0}$ is the exchange stiffness at zero temperature.}
\label{f6}
\end{figure}

\subsection{Exchange stiffness}
The determination of temperature-dependent exchange stiffness constant $A_\text{e}(T)$ is nontrivial. At 300 K, spin-wave dispersion measurements in Nd-Fe-B magnets reveal $A_\text{e}$ as 6.6 pJ/m \cite{ono2014observation}. In the case of uniaxial anisotropy with positive $K_1^\text{eff}$ and zero $K_2^\text{eff}$ and $K_3^\text{eff}$, the domain wall width can be calculated as $\delta_w=\pi \sqrt{A_\text{e}/K_1^\text{eff}}$, from which $A_\text{e}$ is estimated around 7--12 pJ/m at 300 K \cite{durst1986determination,manfred2003micromagnetism}. However, when $K_1^\text{eff}$ is negative or $K_2^\text{eff}$ and $K_3^\text{eff}$ cannot be neglected, e.g. at low temperatures, the expression $\delta_w=\pi \sqrt{A_\text{e}/K_1^\text{eff}}$ does not work. It should be mentioned that if all $K_i^\text{eff}$ are taken into account, there is no analytic solution for the Bloch wall profile \cite{hubert2008magnetic}. In the general case, the Bloch wall profile is governed by \cite{hubert2008magnetic}
\begin{equation}
\text{d}x=\text{d}\theta \sqrt{A_\text{e}(T)/\left[ F(\theta,T)- F(\theta_0,T) \right] }
\label{eq14}
\end{equation}
and thus
\begin{equation}
x(\theta,T)=\sqrt{A_\text{e}(T)} \int_{\theta_0}^\theta \frac{\text{d}\Theta}{\sqrt{F(\Theta,T)- F(\theta_0,T)}},
\label{eq15}
\end{equation}
in which $F(\theta,T)$ is taken from Eq. \ref{eq9}.

Since $x$ is a monotonic function of $\theta$ in Eq. \ref{eq15}, there exists an inverse function $\theta(x,T)$. Therefore, after numerical integration of Eq. \ref{eq15} with various $A_\text{e}(T)$, we attain a series of theoretical curves with $x$ as a function of $M_z=M_s(T)\cos(\theta(x,T))$. Then we optimize $A_\text{e}(T)$ through the least-square method by comparing the simulation data to the theoretical curves. In Fig. \ref{f6}(a), we plot both the simulation data points and the theoretical curves (solid lines) with the optimum $A_\text{e}(T)$. It can be found that the theoretical curves by Eq. \ref{eq15} match well with the fitting results by Eq. \ref{eq12}. But there is intrinsic difference, i.e. Eq. \ref{eq12} only gives domain wall width which can be used to estimate $A_\text{e}$ indirectly through $\delta_w=\pi \sqrt{A_\text{e}/K_1^\text{eff}}$ when $K_1^\text{eff}$ is positive, whereas Eq. \ref{eq15} directly gives $A_\text{e}$ without the constraint on $K_i^\text{eff}$. The optimum $A_\text{e}(T)$ as a function of temperature is presented in Fig. \ref{f6}(b). We can see that $A_\text{e}=(\delta_w/ \pi)^2 K_1^\text{eff}$ yields reasonable results only above 300 K. In general, $A_\text{e}(T)$ shows a decreasing trend as the temperature increases. Below the spin reorientation temperature, $A_\text{e}(T)$ slowly decreases from 11.3 pJ/m at 25 K to 10.2 pJ/m at 140 K. After 200 K, $A_\text{e}(T)$ decreases much faster, from 11 pJ/m at 200 K to 3.5 pJ/m at 575 K.
$A_\text{e}=10.2$ pJ/m at 300 K is also consistent with the literature. However, $A_\text{e}$ decreases more slowly than $K_1^\text{eff}$ with increasing temperature. For instance, from 300 to 500 K, $A_\text{e}$ is reduced by 34$\%$ while $K_1^\text{eff}$ by 85$\%$. This explains the wider domain wall at higher temperature in Fig. \ref{f5}(d).

The scaling behavior of $A_\text{e}(T)$ is presented in Fig. \ref{f6}(c). It is found that at temperatures lower than 500 K, a scaling behavior $A_\text{e}(T) \sim m$ exists. The power exponent of 1 for Nd$_2$Fe$_{14}$B is much lower than 2 in the mean-field approximation (MFA), 1.66 for a simple cubic lattice, and 1.76 for FePt \cite{atxitia2010multiscale}. At temperatures close to $T_\text{c}$, the high-temperature behavior deviates far away from this power scaling law. In addition, fitting the data after 500 K reveals that $A_\text{e}(T)$ approximately follows the scaling law of $m^{1.55}$. Fitting all the data together with low quality gives a scaling law of $m^{1.2}$. The underlying physical reason of this distinct scaling behavior in Nd$_2$Fe$_{14}$B has to be uncovered theoretically in the near future. It should be mentioned that the classical spectral density method has been attempted towards a deep theoretical understanding of the scaling behavior of exchange stiffness for simple cubic, body-centered cubic, and face-centered cubic systems \cite{atxitia2010multiscale,campana1984spectral}, but its application to the complex rare-earth based Nd$_2$Fe$_{14}$B system remains to be further explored.

\section{Conclusions}
In summary, we have carried out \textit{ab-initio} informed atomistic spin model simulations to predict the temperature-dependent intrinsic properties of Nd$_2$Fe$_{14}$B permanent magnets. The results are relevant for temperature-dependent micromagnetic simulations of Nd-Fe-B magnets. The main conclusions are summarized as:

(1) The Hamiltonian of the atomistic spin model for Nd$_2$Fe$_{14}$B includes contributions from the Heisenberg exchange of Fe-Fe and Fe-Nd atomic pairs, the uniaxial single-ion anisotropy energy of Fe atoms, and the crystal-field energy of Nd ions. Specially, we approximately expand the crystal-field Hamiltonian of Nd ions  into an energy formula featured by second, fourth, and sixth-order phenomenological anisotropy constants.

(2) Monte Carlo simulations of the atomistic spin model readily capture the Curie temperature $T_\text{c}$ of Nd$_2$Fe$_{14}$B. After applying the temperature rescaling strategy and the fitted rescaling parameter $\alpha=1.802$, we show the calculated temperature dependence of saturation magnetization $M_\text{s}(T)$ agrees well with the experimental results, and the spin reorientation phenomenon at low temperature is well predicted.

(3) Constrained Monte Carlo simulations give the temperature-dependent total internal torque, from which we calculate the macroscopically effective second-, fourth-, and sixth-order anisotropy constants that match well with the experimental measurements. The calculated values at 300 K shows good consistency with literature reports, with $K_1^\text{eff}$, $K_2^\text{eff}$, and $K_3^\text{eff}$ as 4.26, 0.15, and $-0.10$ MJ/m$^3$, respectively.

(4) Mapping the atomistic magnetic moments to the continuum magnetization leads to the domain wall profile, which can be further fitted by hyperbolic functions to evaluate the domain wall width $\delta_w$. Different domain wall configurations at low temperatures are identified. The calculated $\delta_w$ and its variance increases with temperature, and its value at 300 K is consistent with experimental observation. $\delta_w$ is found to scale with magnetization as a function of $m^{2.26}$.

(5) By using a general continuum formula with the exchange stiffness constant $A_\text{e}(T)$ as a parameter to describe the domain wall profile, we determine $A_\text{e}(T)$. $A_\text{e}$ is found to decrease more slowly than $K_1^\text{eff}$ with increasing temperature. The scaling behavior of the exchange stiffness with the normalized magnetization is found to be $A_\text{e}(T) \sim m$ at temperatures below 500 K and $A_\text{e}(T) \sim m^{1.55}$ at temperatures close to $T_\text{c}$.

\section*{Acknowledgment}
Support from the German Science Foundation (DFG YI 165/1-1 and DFG XU 121/7-1) and the German federal state of Hessen through its excellence programme LOEWE “RESPONSE” is appreciated. M Yi acknowledges the support from the open project of State Key Lab for Strength and Vibration of Mechanical Structures (SV2017-KF-28) and the 15$^\text{th}$ Thousand Youth Talents Program of China. Dr Hongbin Zhang is acknowledged for his valuable comments. The authors also acknowledge the access to the Lichtenberg High Performance Computer of TU Darmstadt.

\bibliography{mybibfile-1}

\end{document}